\documentclass[reprint,superscriptaddress,twocolumn,aps,prb,showpacs,floatfix]{revtex4-1}
\usepackage{graphicx}

\begin{document}

\title{Suppression of magnetic flux avalanches and recovery of the critical state in superconducting NbN films}

\author{V. V. Yurchenko}
\address{Department of Physics, University of Oslo, P.b. 1048, Blindern, Oslo, 0316, Norway}

\author{A. J. Qviller}
\address{Department of Physics, University of Oslo, P.b. 1048, Blindern, Oslo, 0316, Norway}

\author{S. Chaudhuri}
\address{Nanoscience Center, Department of Physics, P.O. Box 35, FI-40014 University of Jyv\"askyl\"a, Finland}

\author{I. J. Maasilta}
\address{Nanoscience Center, Department of Physics, P.O. Box 35, FI-40014 University of Jyv\"askyl\"a, Finland}

\author{D. V. Shantsev}
\address{Department of Physics, University of Oslo, P.b. 1048, Blindern, Oslo, 0316, Norway}

\author{Y. M. Galperin}
\address{Department of Physics, University of Oslo, P.b. 1048, Blindern, Oslo, 0316, Norway}
\address{A. F. Ioffe Physico-Technical Institute RAS, 194021 St. Petersburg, Russian Federation}

\author{T. H. Johansen}
\address{Department of Physics, University of Oslo, P.b. 1048, Blindern, Oslo, 0316, Norway}
\address{Institute for Superconducting and Electronic Materials, University of Wollongong, NSW 2522,
Australia}

\begin{abstract}

Thermo-magnetic instability (TMI) in superconductors is known to destroy
the critical state via magnetic flux avalanches, and hence it deteriorates the ability of the superconductors to shield external magnetic field. In this work, we quantify to what extent the shielding current is affected by TMI. We recover the critical state in one half of a thermo-magnetically unstable NbN film by coating it with a thin layer of Cu. Suppression of the instability in the metal coated part is confirmed by the results of a direct real time magneto-optical imaging.
A pattern of discontinuity lines in the observed flux distribution indicates that only one quarter of the shielding current flows through the whole sample, including the unstable uncoated part, while three quarters flow in loops within the stable Cu-coated part.

\end{abstract}



\maketitle

\section{Introduction}

Thermo-magnetic instability (TMI) in type-II superconductors can emerge in a varying external magnetic field as a result of vortex motion in the superconductor, whereby the moving flux induces electric fields high enough to produce Joule heating larger than heat removal. This suppresses pinning and promotes an avalanche-like influx of a large number of Abrikosov vortices, leading to the destruction of the critical state \cite{Bean64} on a macroscopic level. Magnetic avalanches of this thermo-magnetic origin have been observed in a large family of superconducting films, including Nb,\cite{Wertheimer1967, Aranson} Pb,\cite{Menghini2005} Nb$_3$Sn,\cite{Rudnev2003} NbN,\cite{Rudnev2005, Yurchenko2007, Yurchenko2009} and MgB$_2$.\cite{Johansen2001, JohansenEPL2002} In these films,  the avalanches acquire complex dendritic shapes, as magneto-optical (MO) visualization has revealed.

Because of the dramatic effect that the thermo-magnetic instability (TMI) has on the performance of superconducting devices, it is of primary importance to find ways to suppress it without deteriorating the properties of the material. Due to extensive experimental and theoretical studies of the TMI, there have been considerable advances in understanding its underlying mechanisms.\cite{Mints81, Aranson, Rakhmanov2004, Denisov06, Choi2007, Yurchenko2009} In a series of works on MgB$_2$ films \cite{Baziljevich2002, Choi2009}, it has been demonstrated that the avalanches can be suppressed by coating the superconductor by a few microns thick metal layer.
Although NbN superconducting films are far more widely used in practical devices than those made of MgB$_2$, prospects of suppressing the TMI in NbN films have, to the best of our knowledge, not been investigated. In this work, we study the TMI in a NbN thin film where one half is coated with a layer of Cu. This configuration allows, using MO imaging, direct comparison of coated and uncoated parts without disturbing the intrinsic properties of the superconductor. We show that the metal coating is capable of stabilizing the critical state, and investigate how the current flow is redirected at the interface between the stable and unstable regions.

\section {Experimental}
A film of NbN was prepared and lithographically patterned into a rectangular shape in the following steps. Pulsed laser deposition was used to make the superconducting NbN film on  a single crystalline (100)-oriented MgO substrate at  600$^{\,\circ}$C and nitrogen pressure of $\sim$ 24 mTorr.  More details and deposition parameters can be found elsewhere \cite{Chaudhuri2011}. Then, a 1 $\mu$m thick Cu layer was deposited on top of the NbN using electron-beam (e-beam) evaporation. In the next step, a 400 nm thick layer of e-beam sensitive (positive) PMMA resist was spun on top of the Cu, and  e-beam lithography was carried out such that a rectangular ($2 \times 4$~mm$^2$) region of the polymer  was left unexposed. The exposed polymer (outside the rectangle) was developed 
the Cu and the NbN underneath were etched  chemically and  by reactive ion etching,  respectively  \cite{process},
so that  a rectangular NbN-Cu-PMMA (unexposed) window was left on the substrate, and finally the unexposed polymer was dissolved in acetone. In the final step, once again, a layer of PMMA was spun, and one-half of the rectangular window was exposed and developed, but only the Cu layer underneath was removed this time. In this way, the $2 \times 4$~mm$^2$ NbN sample was covered with Cu on one half of the area.

Details of the flux propagation were visualized using MO imaging based on the Faraday effect in Bi-substituted yttrium-iron garnet films \cite{Goa2003, Helseth2002}. The NbN sample was mounted on the cold finger of an optical He-flow cryostat from Oxford Instruments and observed using a Leica polarized light microscope. A magnetic field applied perpendicular to the sample was controlled by a pair of resistive coils. In our experimental setting, the brightness of the MO images is proportional to the square of the local flux density. The images were recorded by a 12-bit high sensitivity CCD camera with linear response, Retiga-Exi from QImaging. Field control and image acquisition were automated and synchronized using the National Instruments LabView programming environment. A series of zero-field-cooled experiments carried out at different temperatures, while sweeping the magnetic field from zero to the fully penetrated state, i.e., when the flux front reaches the center of the sample, and back to zero, to the so-called remanent state.

\section{Results and Discussion}

\subsection{Stable Regime}

\begin{figure}[t]
\centering
  \includegraphics*[width=0.9\columnwidth]
  {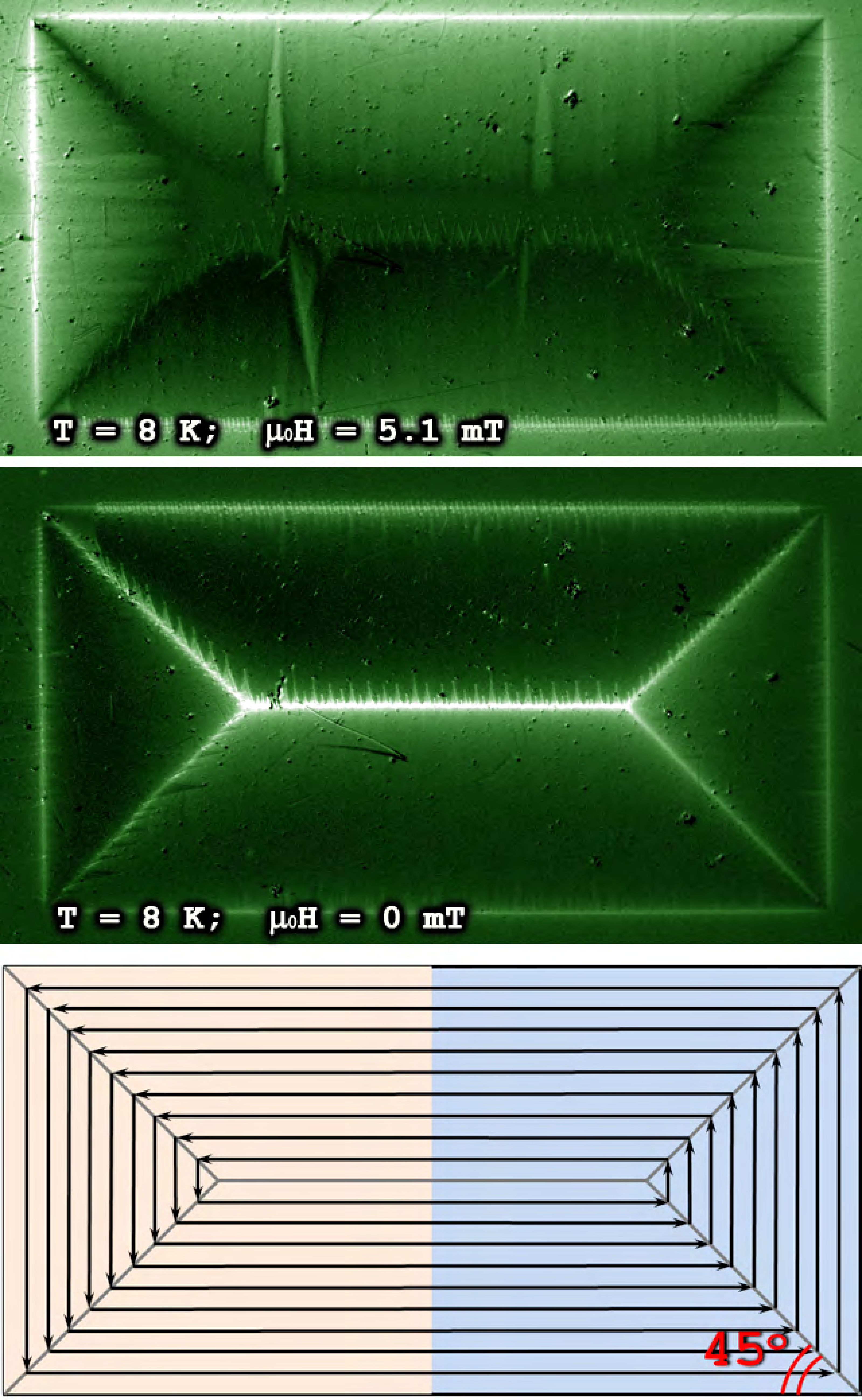}
  \caption{ (Color online) MO images of NbN films with the left half coated with Cu at T = 8 K in ascending magnetic fields (top) and in the remanent state (middle). Schematic drawing of the current stream lines explaining appearance of the d-lines (bottom).}
  \label{MO8K}
\end{figure}
One of the characteristic features of TMI is that it
exists only below a certain threshold temperature $T_{\text{th}}$.  Above this temperature, which in our case was $T_{\text{th}} \approx 7.1$~K, the flux penetrated gradually and the flux distribution
showed the critical state behavior. This is demonstrated in
Fig.~\ref{MO8K}, displaying MO images taken at 8~K in increasing magnetic field (top panel) and in the remanent state (middle panel).
Note that above $T_{th}$ there was no difference in the flux propagation in the coated and the uncoated halves,
implying that the metal layer had no effect on the flux distribution in the stable regime. It indicates that the lithographic processing did not introduce any detectable inhomogeneities in the sample.

The remanent state displayed a characteristic pattern of d-lines -- current discontinuity lines -- that are
seen as bright lines in the MO image. D-lines are a geometric set of points where at least one component of the current
changes abruptly.\cite{Jooss2002}  The relatively simple pattern in the middle panel of Fig.~\ref{MO8K} can be understood from a straightforward geometric construction presented in the bottom panel of the figure. It shows schematically the current stream lines that are drawn with a constant spacing inversely proportional to the critical current density $j_c$. In this simple case, the d-lines are straight and the angle between them and the sample edges is 45 degrees.

\subsection{Instability and Recovery of the Critical State}

\begin{figure}[ht]
\centering
  \includegraphics*[width=0.9\columnwidth]
  {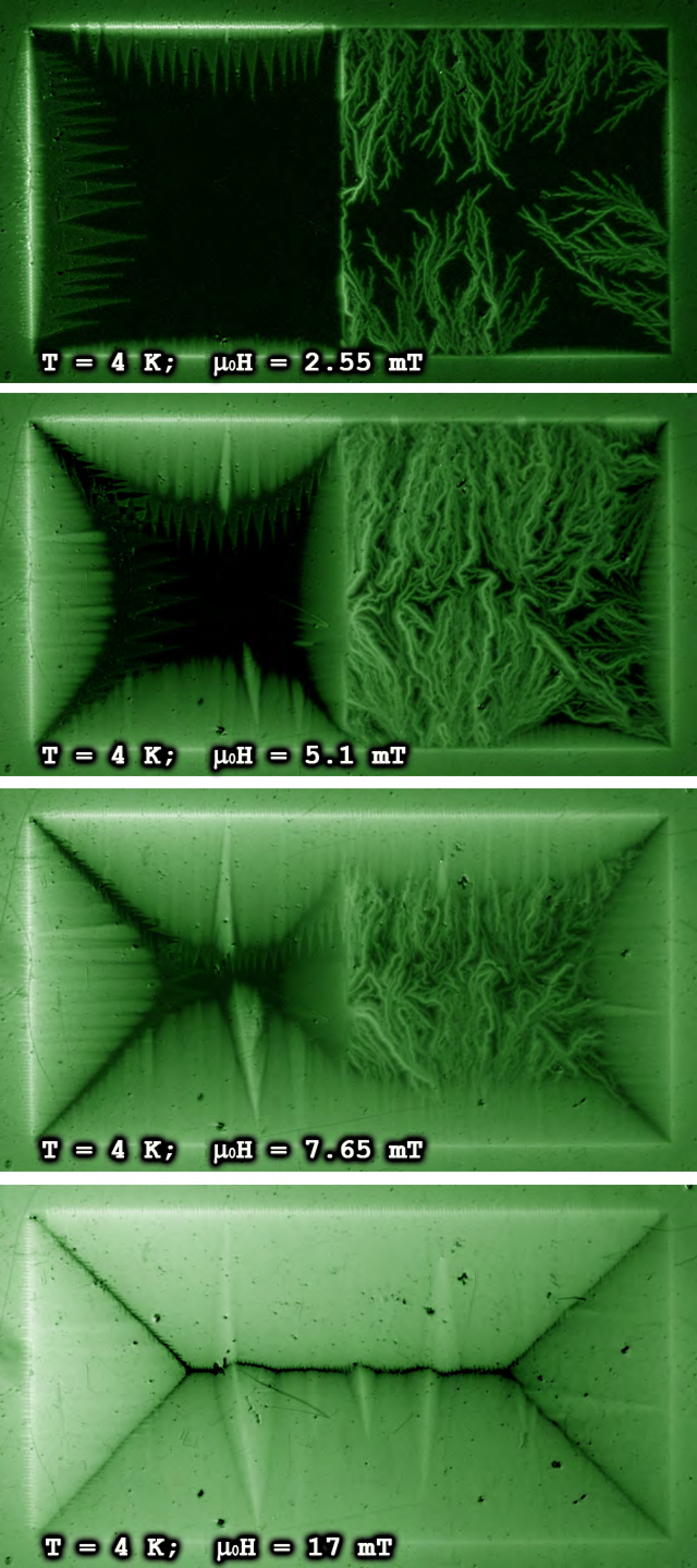}
  \caption{ (Color online) MO images of NbN superconducting film, left half of which was coated with $1~\mu$m thick layer of Cu, in ascending magnetic fields after zero-field-cooling to $T = 4$~K.}
  \label{MOAsc4K}
\end{figure}
\begin{figure}[ht]
\centering
  \includegraphics*[width=0.9\columnwidth]
  {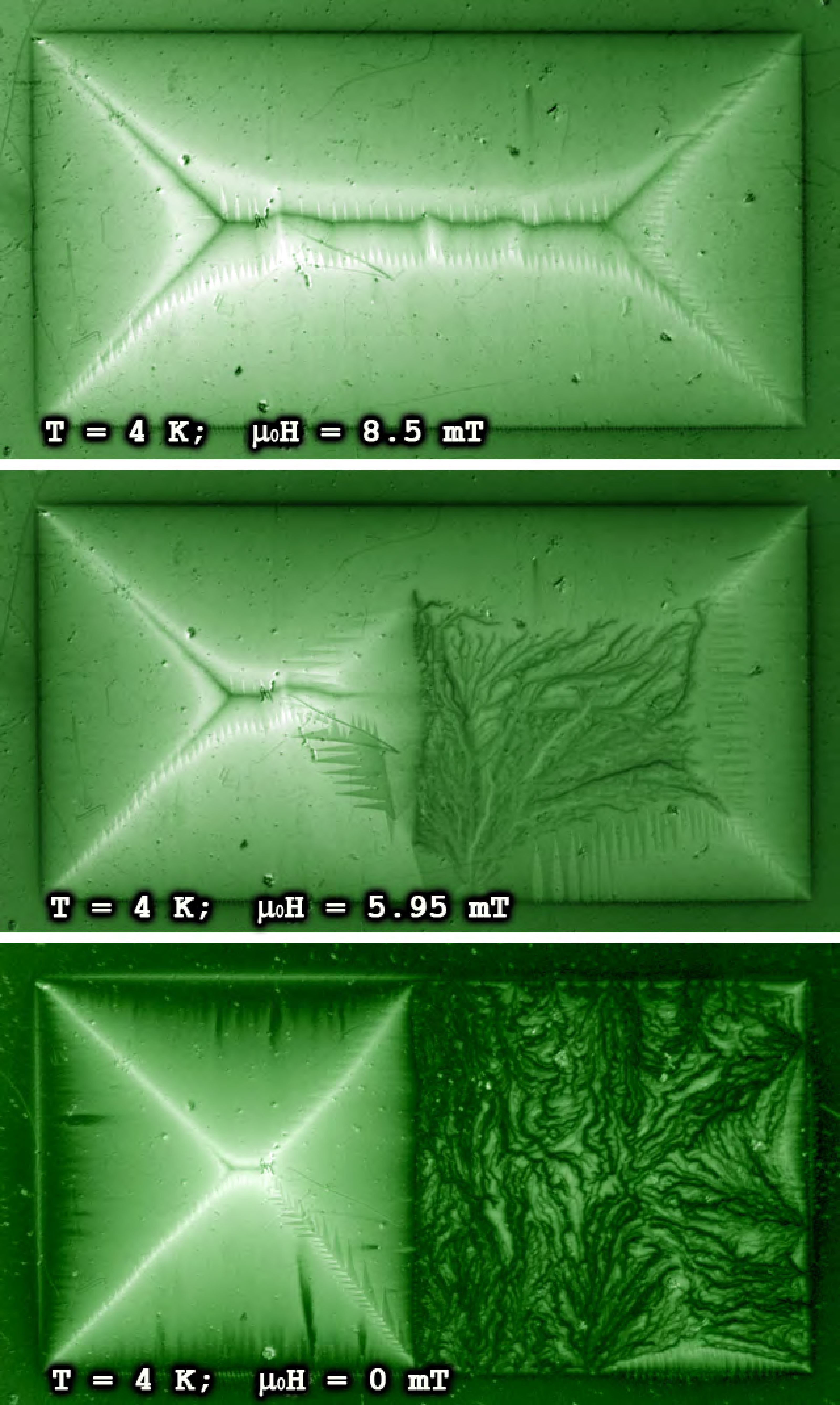}
  \caption{ (Color online) MO images of NbN superconducting film, left half of which was coated with Cu, in descending magnetic fields at $T = 4$~K.}
  \label{MODesc4K}
\end{figure}

Below $T_{\text{th}}$, the NbN film entered the instability regime.
Shown in  Fig.~\ref{MOAsc4K} are MO images of the  NbN
sample in ascending magnetic fields at $T=4$~K.
The instability is known to exist only within a certain interval of the external magnetic field,
between the lower $H_1^{\text{th}}$ and the upper $H_2^{\text{th}}$ threshold fields.\cite{Yurchenko2007}
In the uncoated part of the sample the first avalanche was observed at $\mu_0 H_{1}^{\text{th}}(4 \text{K}) \approx 0.1$~mT.
In the magnetic field of $\mu_0 H = 2.55$ ~mT dendritic flux avalanches invaded most of its volume, while no traces of TMI were found under the Cu layer.
Metallization of the NbN film affected avalanches at every stage of their development: in addition to the fact 
that their nucleation was totally suppressed in the NbN/Cu part, a complete screening was also observed, i.e. even those dendrites that appeared in the bare part and propagated towards the left half could not penetrate under the metal.
As can be seen in the MO images, all the branches of the incoming avalanches that reached the border with the Cu covered region changed direction and traveled along it.

As the field was increased further, an additional flux front began to form along the border between the two parts,
as if they were strongly electromagnetically decoupled, with the critical state on the left hand side, and
essentially a disordered flux distribution on the right side.
The central flux front kept on growing into the Cu-covered part until the upper threshold field $\mu_0 H_2^{\text{th}} \approx 6$~mT was reached, above which the critical state started to recover everywhere in the sample and the flux front gradually acquired its usual shape.
Disappearance of the additional vertical flux front above $H_2^{\text{th}}$ and onset of a \textit{regular} critical state at full penetration (see, e.g.,
the panel corresponding to  $\mu_0 H=17$~mT in Fig.~\ref{MOAsc4K})  prove that the observed features are not artifacts that appear often as a result of sample degradation during lithographic treatment. That was also confirmed by the visualizations at $T = 8$~K (Fig. \ref{MO8K}) and at $T = 4$~K in descending magnetic fields $H > H_2^{\text{th}} $.
Upon reaching the maximum value of $\mu_0 H = 17$~mT, the field was gradually reduced. In the interval of descending fields $\mu_0 H \in [ 17; 5.95)$~mT a uniform negative flux front was smoothly penetrating the sample (Fig.~\ref{MOAsc4K}). At 5.95~mT the first dendritic avalanche with negative polarity invaded the right half.
At lower fields numerous avalanches completely destroyed the critical state there and in the remanent state the flux in the uncoated part
was distributed in quite a chaotic manner (bottom panel).
At the same time, under the Cu layer a critical state without any signs of instability was formed.
The flux distribution there had a characteristic dome shape similar to that observed in the stable regime (Fig. \ref{MO8K}).

\subsection{Evaluation of the shielding current reduction due to TMI }

A methodological beauty of MO visualization is that it allows quantitative assessments based on a straightforward geometrical
analysis of the images. For instance, analysis of the d-lines permits a reconstruction of the current flow profiles.
In SCs with heterogeneous  critical current densities, d-lines mark the borders of domains with different directions and/or values of $j_c$ and may serve for evaluation of inter-granular currents\cite{WelpDline1994, PolyanskiiDLine1996}, anisotropy parameters,\cite{Yurchenko2010} etc.

In this regard, the  pattern of d-lines shown in the bottom panel of Fig.\ref{MODesc4K} is particularly interesting.
In the remanent state, the bare NbN part (right) was strongly affected by TMI due to a very high avalanche activity in descending fields. Despite a chaotic flux distribution in the uncoated part, in the NbN/Cu half (left) the flux distribution acquired a typical dome-shape. It only slightly deviated from the one in an isotropic square sample, where the angles between the d-lines and the edges are exactly $45^{\circ}$. However, in our case, due to the shielding currents flowing in the uncoated part, the angle $\alpha$ between the central d-line and the sample edge was larger than $45^{\circ}$. By measuring $\alpha$, one can quantify the ratio of the effective critical current density in the coated part, $j_c$, to the current density leaking to the uncoated part,  $j_{\text{TMI}}$.  We use the approach developed for evaluation of a current across a weak-link in superconducting films on bi-crystal substrates, \cite{WelpDline1994, PolyanskiiDLine1996} which gives
\begin{equation}
j_{\text{TMI}}/j_c = - \cos 2 \alpha.
\label{ratio}
\end{equation}

\begin{figure}[t]
\centering
  \includegraphics*[width=0.9\columnwidth]
  {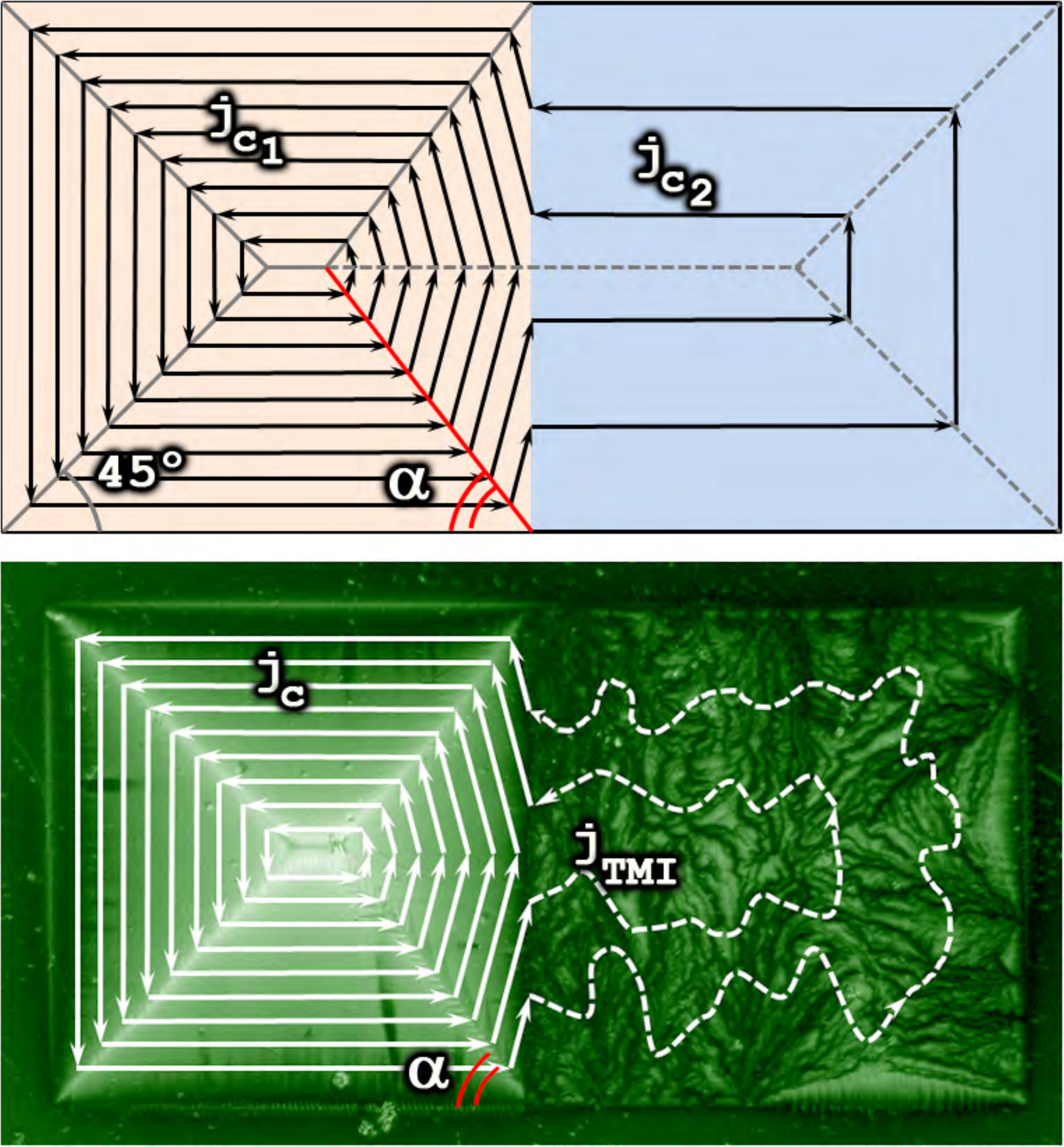}
  \caption{ (Color online) (Top) Schematic reconstruction of the current flow profile in an idealized SC sample where the critical current density in the left part is factor four higher than across the border. (Bottom) Schematic representation of the current stream lines in the actual sample.}
  \label{DlinesSchematic}
\end{figure}

At temperature $T=4$~K, $\alpha  \approx 51.9 \pm 0.6^{\circ}$, which means that the instability reduced the current approximately by a factor of four compared to its critical value determined by the pinning alone:
 $j_{\text{TMI}}/j_c\approx (0.24 \pm 0.02)$.
Since the remanent state is the one most affected by the instability, the value obtained for the ratio of the currents $ \approx 4 $ gives the
upper estimate for the effective critical current enhancement due to the suppression of TMI.
Figure~\ref{DlinesSchematic} schematically shows a reconstruction of the current flow profiles.
The upper panel shows the critical state in a hypothetical sample with two $j_c$'s that differ by factor four: $j_{c_1} = 4 \cdot j_{c_2}$. The d-lines that stem from such a current distribution in the center of the sample form an angle $\alpha \approx 52^{\circ}$ with the edges.
Apparently, the real situation, schematically shown in the bottom panel, was more complex. The current distribution in the uncoated part of the sample was strongly inhomogeneous due to the stochastic nature of the avalanches. However, absence of specific features on the vertical flux front in the center of the sample indicates that the current was distributed fairly uniformly along the border. The bottom panel in Fig. \ref{DlinesSchematic} shows very schematically the current flow profiles in the actual situation. The current stream lines in the left half are drawn with a constant distance $\sim 1/j_c$, while the spacing between the current lines along the border with the right half is increased by factor four $ \sim 4/j_c$, in accordance with the experimental picture.

\begin{figure}[t]
\centering
  \includegraphics*[width=0.9\columnwidth]
  {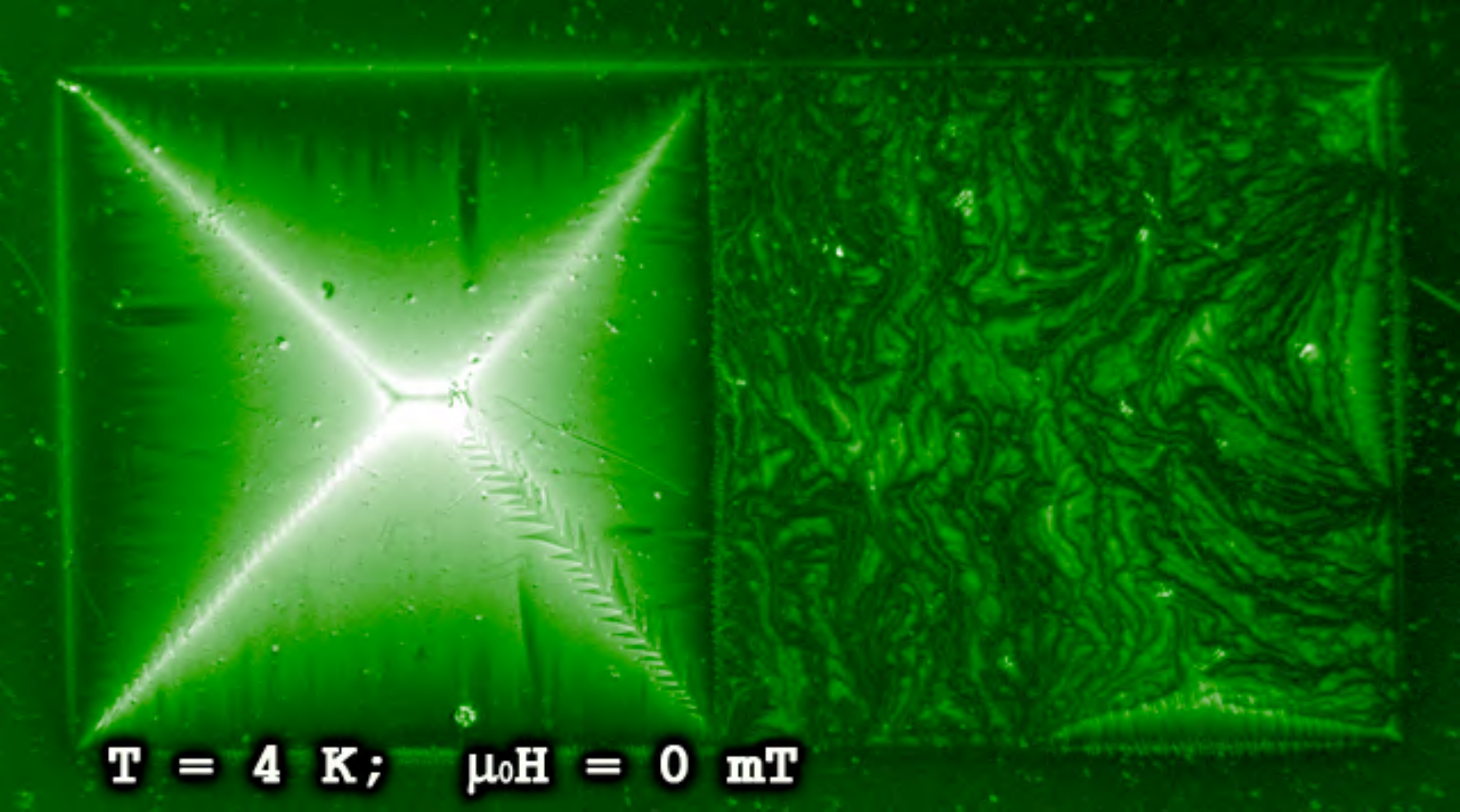}
  \caption{ (Color online) MO image of NbN superconducting film, left half of which was coated with Cu, in the remanent state upon field cycling at $T = 4$~K. Despite irreproducible chaotic flux distribution in the right half, the critical state in the Cu-coated part shows exactly the same d-line pattern.}
  \label{Reprod}
\end{figure}

Also remarkable is that the angle $\alpha$ was quite robust. This was checked by cycling the magnetic field and remeasuring the angle $\alpha$ in the remanent state. The MO image presented in Fig. \ref{Reprod} was taken upon reversing the field polarity, reaching the maximum negative field and reducing it again to zero. The exact details of the remanent state in the unstable half were irreproducible. At the same time, the critical state in the Cu-coated part was well reproduced. Namely, the d-line pattern was preserved, which means that the ratio of the currents $j_{TMI} / j_c$ remained constant.

In conclusion, we studied thermo-magnetic instability in NbN films and found that the magnetic flux avalanches can be effectively suppressed by an overlay Cu layer without any damage to the superconducting film itself. The avalanches were not only suppressed by the metal coating, but also completely screened, resulting in a perfect recovery of the critical state and enhancement of the shielding current density approximately by factor $4$ compared to the part of the film with TMI.

\begin{acknowledgments}
This work was in part supported by the Norwegian Research Council. Fruitful discussions with Dr. A. Polyanskii and Dr. F. Laviano are acknowledged. Contribution of Dr. E. Ilyashenko and Dr. A. Solovyev during the preparation of MO indicators is greatly appreciated. Research  at  University of Jyv\"askyl\"a was supported by  the Academy of Finland project number 128532.
\end{acknowledgments}


\begin{thebibliography}{24}
\bibitem [{pro()}]{process}%
  \BibitemOpen
  \href@noop {} {}\bibinfo {note} {The exposed resist was dissolved using a
  solution of methyl-isobutyl-ketone and isopropanol in a volume ratio of 2:1.
  The copper was etched using a solution of acetic acid, hydrogen peroxide and
  de-ionised water. The NbN was etched using a mixture of CF$ _{4} $ and O$
  _{2} $ gas.}\BibitemShut {Stop}%
\end{thebibliography}
%

\end{document}